\documentclass[12pt]{iopart}

\usepackage{iopams}
\usepackage[mathlines]{lineno}
\usepackage{delarray}
\usepackage{epsfig}
\usepackage{epstopdf}
\usepackage{color}
\usepackage{mathrsfs}

\usepackage{graphicx}% Include figure files
\usepackage{dcolumn}% Align table columns on decimal point
\usepackage{bm}% bold math
\usepackage[mathlines]{lineno}% Enable numbering of text and display math
\usepackage{multirow}
\newcommand{\rr} {\boldsymbol{r}}

\begin{document}

\title[Phys. Scr]{Towards microscopic studies of survival probabilities of compound superheavy nuclei}

\author{Yi Zhu \& J C Pei}

\address{State Key Laboratory of Nuclear
Physics and Technology, School of Physics, Peking University,  Beijing 100871, China}
\ead{peij@pku.edu.cn}
\vspace{10pt}
\begin{indented}
\item[]May 2017
\end{indented}

\begin{abstract}
The microscopic approach of fission rates and neutron emission rates in compound nuclei have been applied
to $^{258}$No and $^{286}$Cn. The microscopic framework is based on the finite-temperature
Skyrme-Hartree-Fock+BCS calculations, in which the fission barriers and mass parameters are self-consistently
temperature dependent. The fission rates from low to high temperatures can be obtained
based on the imaginary free energy method. The neutron emission rates are obtained
with neutron gases at surfaces. Finally the survival probabilities of superheavy nuclei can be calculated microscopically.
 The microscopic approach has been compared with the widely used
statistical models. Generally, there are still large uncertainties in descriptions of fission rates.

\end{abstract}
\vspace{2pc}

\date
\maketitle
% Uncomment for PACS numbers
%\pacs{00.00, 20.00, 42.10}
%
% Uncomment for keywords
%\vspace{2pc}
%\noindent{\it Keywords}: XXXXXX, YYYYYYYY, ZZZZZZZZZ
%
% Uncomment for Submitted to journal title message
%\submitto{\JPA}
%
% Uncomment if a separate title page is required
%\maketitle
%
% For two-column output uncomment the next line and choose [10pt] rather than [12pt] in the \documentclass declaration
%\ioptwocol
%

\ioptwocol

\section{Introduction}

To quest the heaviest nuclei, whose existences are merely due to quantum shell effects,  is one of the major issues in nuclear physics~\cite{oganessian2017,hofmann2000,itkis2015}.
Very recently, four elements with Z=113, 115, 117 and 118 were officially named as Nihonium (Nh), Moscovium (Mc), Tennessine (Ts), Oganesson (Og), respectively by IUPAC~\cite{ohrstrom2016}.
Up to date, nuclei with proton numbers up to 118 have been experimentally discovered and confirmed.
There are typical cold~\cite{armbruster1985,hofmann2002} and hot fusion~\cite{oganessian2010} reactions to synthesize superheavy nuclei.
The key question is to find the optimal combination of beam-target and the bombarding energy in order to maximize the production cross sections.
It will be a much harder challenge to synthesize new elements beyond Z=118.

The synthesis procedure of superheavy nuclei can be described as the capture-fusion-evaporation reaction.
The final production cross section (or evaporation residue cross section) can be written as~\cite{itkis2015}
\begin{equation}
\sigma_{EVR}= \sum_J \sigma_c^J(E_{\mathrm{c.m.}}) P^J_{\mathrm{CN}}(E^{*}) W^J_{\mathrm{sur}}(E^{*} )
\label{eq1}
\end{equation}
 In Eq.(\ref{eq1}), $\sigma_c$ is the capture cross-section,
$P_{\mathrm{CN}}$ is the fusion probability of the compound nuclei, $W_{sur}$ is the survival probability
of the compound nuclei. The survival probability is mainly determined by the competition between
the neutron emission rates and fission rates in compound nuclei.
The $\alpha$ decays can provide critical information of ground states of superheavy nuclei but are negligible
 compared to the rapid fission and neutron emission processes
 in highly-excited compound superheavy nuclei.
 Generally, there are large uncertainties in theoretical descriptions of these three steps although
the total production cross section can be reproduced by various parameterized models. Experimentally,
the measurement of survival probabilities is feasible. For example, very large survival probabilities of $^{258}$No~\cite{peterson2009} and
$^{274}$Hs~\cite{Yanez2014} have been directly obtained in hot fusion reactions, which provide a good opportunity to verify various theoretical models.

Conventionally, the statistical models have been widely applied to the calculations of survival probabilities $W_{\mathrm{sur}}$
of highly excited nuclei~\cite{zubov2002,zubov2005,xiacj2011}. The pioneer applications of statistical model can be traced back to Weisskopf for neutron evaporation in 1937~\cite{weisskopf}
and Bohr-Wheeler for fission in 1939~\cite{bohr1939}.
The statistical model of fission, also called transition state theory, involves fission barriers and level densities at ground state and saddle point, respectively.
There are many developments on the statistical models with adjusted parameters and collective corrections.
In particular, whether the fission barriers and level density parameters are temperature (or excitation energy) dependent is still a question~\cite{swiaatecki,adamian2010}.
On the other hand, the microscopic descriptions of survival probabilities are based on effective nuclear forces and
there are non-adjusted parameters needed~\cite{schunck2016}.
The nuclear density functional theory is an ideal theoretical tool for descriptions of heavy and superheavy nuclei.
The microscopic fission theory based on finite-temperature nuclear density functional theory
can self-consistently describe the thermal properties of compound nuclei and the gradually decreased quantum effects.
It is still worth to understand the microscopic fission mechanism~\cite{schunck2016} so as to make predictions for unknown experiments, although
phenomenological statistical models have been widely used.

In this work, we introduce the microscopic framework for descriptions of the fission rates~\cite{zhu2016},
neutron emission rates~\cite{zhu2014} and then survival probabilities based on the finite-temperature
Skyrme Hartree-Fock-Bogliubov (or BCS) theory~\cite{goodman1981}. In our approach,
the fission barriers are given in terms of free energies and are temperature dependent~\cite{sheikh2009}.
The collective inertia mass parameters are calculated with the temperature dependent cranking approximation~\cite{iwamoto1979,baran1994}.
Then the fission rates are obtained with the imaginary free energy (IMF) method~\cite{langer1967,affleck1981} from low to high temperatures.
The HFB solutions in coordinate spaces can self-consistently produce neutron gases around surfaces~\cite{pei2010}.
Then the neutron emission rates can be related to the neutron gas density~\cite{zhu2014}.
For comparison, we also studied the survival probabilities with the widely used statistical models.

This paper is organized as follows. In Sec.\ref{theory}, we review the two theoretical methods to calculate the survival probability of the compound superheavy nuclei.
 In Sec.\ref{result}, our results of $^{258}$No and $^{286}$Cn are presented  and compared with experimental data.
The summary and our perspectives are given in Sec.\ref{summary}.

\section{Theoretical framework}\label{theory}

\subsection{FT-HFB}

The finite-temperature Hartree-Fock-Bogoliubov (FT-HFB) theory was firstly derived by Goodman in 1981~\cite{goodman1981}. We only display some relevant equations here.
The FT-HFB equation in the coordinate space is written as~\cite{khan2007}:

\begin{equation}
  \left[
\begin{array}{cccc}%
 \displaystyle h_{T}-\lambda& {\hspace{0.7cm} } \Delta_{T} \vspace{2pt}\\
  \displaystyle \Delta_{T}& -h_{T}+\lambda \\
\end{array}
\right]\left[
\begin{array}{clrr}%
u_i \\ \vspace{2pt} v_i\\
\end{array}
\right]=E_i\left[
\begin{array}{clrr}%
 u_i  \vspace{2pt}\\ v_i
\end{array}
\right],
\label{HFB}
\end{equation}
where $h_{T}(\rr)$ and $\Delta_{T}(\rr)$ are the temperature-dependent single-particle Hamiltonian and pairing potential, respectively.
For the particle-hole interaction channel, the SkM* interaction\cite{bartel1982} is employed.
The density-dependent pairing interaction\cite{dobaczewski2002} is adopted in the particle-particle channel.
The FT-HFB equation has the same form with the HFB equation at zero temperature,
but the density $\rho(\rr)$ and pairing density $\tilde{\rho}(\rr)$ are modified as

\begin{eqnarray}
\rho(\rr)=\sum_i|u_i(\rr)|^2f_i+|v_i(\rr)|^2(1-f_i),\\
\tilde{\rho}(\rr)=\sum_iv_i(\rr)^{*}(1-2f_i)u_i(\rr),
\label{rho}
\end{eqnarray}
where the temperature dependent factor $f_i$ is
\begin{equation}
f_{i}=\frac{1}{(1+e^{E_{i}/kT})}
\end{equation}
The entropy $S$ is evaluated with the finite temperature HFB appraximation as~\cite{goodman1981}:
\begin{equation}
S=-k\sum_{i}[f_i\mathrm{ln}f_i+(1-f_i)\mathrm{ln}(1-f_i)].
\end{equation}
At a constant temperature $T$, the free energy is given as $F = E - TS$.
We used the HFB-AX solver~\cite{pei2008} with finite temperatures in deformed coordinate spaces to study neutron emission rates.
The details of calculations can be found in the previous paper~\cite{zhu2014}. The finite-temperature Hartree-Fock+BCS equation
can be solved similarly, which is computationally more efficient for thermal fission studies~\cite{zhu2016}.

The essential inputs for the fission studies includes the fission barriers and mass parameters.
The fission barriers are given in free energies. The fission barriers are self-consistently temperature
dependent, including the fission barrier heights and the barrier curvatures. The mass parameters as a function of deformation $\beta_{20}$
are calculated by the temperature dependent cranking approximation~\cite{iwamoto1979,baran1994}, as written as
\begin{equation}
 M_{20} = \hbar^2 [\mathcal{M}^{(1)}]^{-1}[\mathcal{M}^{(3)}][\mathcal{M}^{(1)}]^{-1}
\end{equation}
 \begin{equation}
 \begin{array}{l}
\mathcal{M}_{ij, T}^{(K)} =\displaystyle \frac{1}{2}\sum<0|Q_i|\mu\nu><\mu\nu|Q_j|0> \vspace{5pt}\\
              \displaystyle \Big\{ \frac{(u_\mu u_\nu - v_\mu v_\nu)^2}{(E_\mu-E_\nu)^K}\big[\tanh(\frac{E_\mu}{2kT})-\tanh(\frac{E_\nu}{2kT})\big]  \vspace{5pt}\\
              +\displaystyle \frac{(u_\mu v_\nu + u_\nu v_\mu)^2}{(E_\mu+E_\nu)^K}\big[\tanh(\frac{E_\mu}{2kT})+\tanh(\frac{E_\nu}{2kT})\big] \Big\}  \vspace{5pt} \\
\end{array}
\end{equation}
where $v_{\mu}^2$ is the BCS occupation number; $E_{\mu}$ is the BCS quasiparticle energy.

\subsection{neutron emission rates}

In the coordinate-space FT-HFB calculation, the external neutron gas is produced naturally.
The neutron emission width $\Gamma_n$ of the compound nucleus is given by the nucleosynthesis formula~\cite{bonche1984}:
\begin{equation}
\frac{\Gamma_n}{\hbar} = n<\sigma v>
\end{equation}
 where $\sigma$ is the neutron capture cross section defined as $\pi R^2$, $n$ denotes the neutron gas density, and $v$ is the average velocity of the external gas~\cite{zhu2014}.
 The calculations of neutron emission rates don't involve level densities and free parameters, see details in Ref.~\cite{zhu2014}.

\subsection{nuclear fission rates}

The WKB method has been widely used for descriptions of the spontaneous fission lifetime~\cite{baran2011,erler2012}.
There are two key inputs, the fission barriers and the collective mass parameters, for the WKB calculations.
It is known that the SkM* force~\cite{bartel1982} can give reasonable fission barriers  and has widely been used for fission studies. The mass parameters can
be calculated by the cranking approximation and the temperature-dependent cranking approximation~\cite{iwamoto1979}.
For thermal excited nuclei, the fission rates can be estimated by the imaginary free energy method~\cite{affleck1981,hagino1996}.
At low temperatures, the fission is mainly the barrier tunneling process.
While at high temperatures, the fission is basically the barrier reflection process.
The general IMF formula\cite{miller1975,hanggi1990} for the decay rates from excited systems is given as:
\begin{equation}
\Gamma =  Z^{-1}\int_0^{\infty}dE P(E)\mathrm{exp}(-\beta E )
\label{Gamma}
\end{equation}
where $P(E)$ is related to the transmission probability.
$Z$ indicates the normalization factor, and it is actually the partition function in the metastable system.
The above formula applies to quantum systems in an ideal heating bath, which is suitable for chemical reactions but
is not a good approximation for nuclear reactions. In this case, the integral upper limit may be modified
to the excitation energy $E^*$ to be consistent with the statistical model.

The temperature-dependent potential valley around the metastable equilibrium deformation can be approximated to be a harmonic oscillator well.
Then the partition function can be derived as~\cite{affleck1981}:
\begin{equation}
Z =  \sum_{n=0}^{\infty}e^{-(n+\frac{1}{2})\hbar\omega_0\beta}
  = [2 \mathrm{sinh}(\frac{1}{2}\beta\hbar\omega_0)]^{-1}
  \label{Z}
\end{equation}
where $\beta = 1/kT$ , $\omega_0$ is the curvature of the potential well.
For realistic potential barriers and mass parameters, we can extract $\omega_0$ approximately by~\cite{zhu2016}:
\begin{equation}
\omega_0 = \pi E / \int_a^b \sqrt{2M(s)(E-V(s))}ds
\end{equation}
$M(s)$ is the temperature-dependent mass parameter along the fission path,
and it can be estimated by the temperature-dependent cranking approximation~\cite{iwamoto1979,baran1994}.

At low temperatures, the fission probability at $E^*$ is given by the WKB method as,
\begin{equation}
\displaystyle P(E)=  e^{ -\frac{2}{\hbar}\int_b^c ds\sqrt{2M(s)(V(s)-E)}}
\label{pe}
\end{equation}
By combining Eqs.(\ref{Gamma}, \ref{Z} and \ref{pe}), the averaged low-temperature fission rates can be obtained.

 We can approximate the temperature-dependent barrier as an inverted harmonic oscillator potential. The barrier curvature
 $\omega_b$ at the saddle point can be extracted by~\cite{zhu2016},
 \begin{equation}
\omega_b = \pi(V_b-E) / \int_b^c \sqrt{2M(s)(V(s)-E)}ds
\end{equation}
where $V_b$ denotes  the barrier height.

For fission rates at high temperatures, the contribution is dominated by reflections above the barriers.
In this case, the fission probability  $P(E)$ can be estimated by,
\begin{equation}
P(E) = (1+\mathrm{exp}[2\pi (V-E)/\hbar\omega_b])^{-1}
\end{equation}
Finally the averaged fission rates at high temperatures can be written as~\cite{affleck1981},
\begin{equation}
\Gamma_f = \frac{\omega_b}{2\pi}\frac{\sinh(\frac{1}{2}\beta\omega_0)}{\sin(\frac{1}{2}\beta\hbar\omega_b)}\exp(-\beta V_b),
\label{espon3}
\end{equation}

\subsection{Statistical model}

Statistical models have been  widely used for calculating the survival probabilities of superheavy nuclei~\cite{zubov2002,xiacj2011}.
In the statistical model, the width of neutron evaporation is:

\begin{equation}
\begin{array}{ll}
\Gamma_n(E^*)=&\frac{2mR^2}{\pi \hbar^2\rho(E^*)}  \vspace{5pt} \\
  &\times\int_{0}^{E^{*}-B_n}\varepsilon_n \rho(E^{*}-B_n-\varepsilon_n) d\varepsilon_n \\
  \end{array}
\label{stat}
\end{equation}
where $m$ is the neutron mass; $R$ is the radius of compound nucleus; $B_n$ is the neutron separation energy; $\rho(E^*)$ is the level density at energy $E^*$ of the ground-state deformation.

 The fission width is calculated by the Bohr-Wheeler formula~\cite{bohr1939}:
\begin{equation}
\begin{array}{l}
\Gamma_f(E^*) =  \displaystyle \frac{1}{2\pi \rho(E^*)} \times \vspace{5pt}\\
~~~~  \int_{0}^{E^{*}-B_f} \rho_{s.d.}(E^{*}-B_f-\varepsilon_f)T_f(\varepsilon_f)  d\varepsilon_f, \\
\end{array}
\label{stat}
\end{equation}
where $B_f$ is the fission barrier; $\rho_{s.d.}$ is the level density at the saddle point.
 The fission barrier height $B_f$ is conventionally
taken as the ground state barrier. However, the temperature or excitation energy dependent barrier
may be more reasonable.
The barrier transmission probability $T_f(\varepsilon_f)$ is defined as:
 \begin{equation}
 T_f(\varepsilon_f) = \Big\{ 1+\mathrm{exp} [-\frac{2\pi\varepsilon_f}{\hbar\omega_{s.d.}}] \Big\} ^{-1}
 \end{equation}
the curvature is $\hbar\omega_{s.d.}=2.2$ MeV, as suggested in Refs.~\cite{zubov2005,xiacj2011}.

Usually the level density is calculated by the Fermi-gas model with several corrections according to Ref.~\cite{xiacj2011},
 \begin{equation}
 \frac{ \rho(E^*) }{2J+1}   = \frac{\mathrm{exp}[2\sqrt{a(E^*-\delta)}-\frac{(J+1/2)^2}{2\sigma^2}]}{24\sqrt{2}\sigma^3a^{1/4}(E^*-\delta)^{5/4}}
 \end{equation}
with
  \begin{equation}
  \sigma^2 = 6\bar{m}^2\sqrt{a(E^*-\delta)}/\pi^2 ,\quad \bar{m}^2 \approx0.24A^{2/3}
  \end{equation}
 In this work, the level density parameter is  $a=A/12 $MeV$^{-1}$. The level density parameter at the saddle point is $a_{s.d.}=1.1A/12 $MeV$^{-1}$.
 The pairing correction adopts $\delta = 12 /\sqrt{A}$ for even-even nuclei.

\section{Results and Discussions}\label{result}

In this section, we study the survival probabilities of  $^{258}$No and $^{286}$Cn,
for which the experimental data are available.  Therefore we can comparatively analyze the microscopic approach and the statistical model in details.

\begin{figure}[t]
  % Requires \usepackage{graphicx}
  \includegraphics[width=0.48\textwidth]{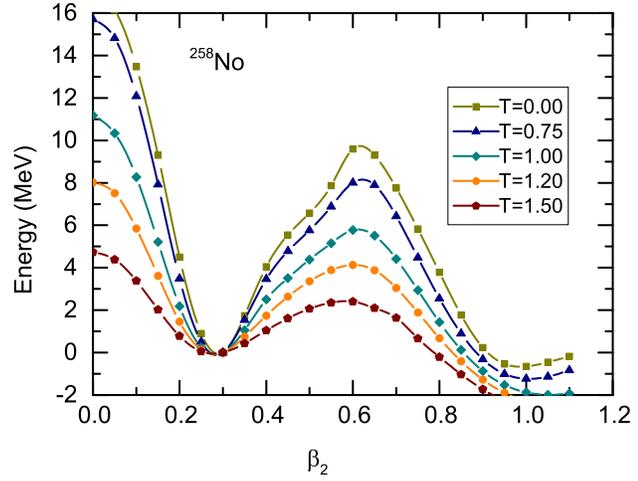}\\
  \caption{(Color online) Calculated temperature dependent fission barriers of $^{258}$No  as a function of
  quadrupole deformation $\beta_2$. The unit of the temperature is MeV.   }
  \label{fig-barrier}
\end{figure}

Fig.\ref{fig-barrier} shows the temperature dependent fission barriers of $^{258}$No in axial-symmetric calculations,
using the coordinate-space solver Skyax~\cite{reinhard}.
Note that the non-axial deformation may decrease the fission barrier height.
We can see that the fission barriers gradually decrease with increasing temperatures.
This is consistent with the fact that shell effects disappear with increasing temperatures~\cite{egido2000}.
Note that the damping factors of shell effects are dependent on proton/neutron numbers in microscopic calculations~\cite{sheikh2009,pei2009}.

\begin{figure}[t]
  % Requires \usepackage{graphicx}
  \includegraphics[width=0.48\textwidth]{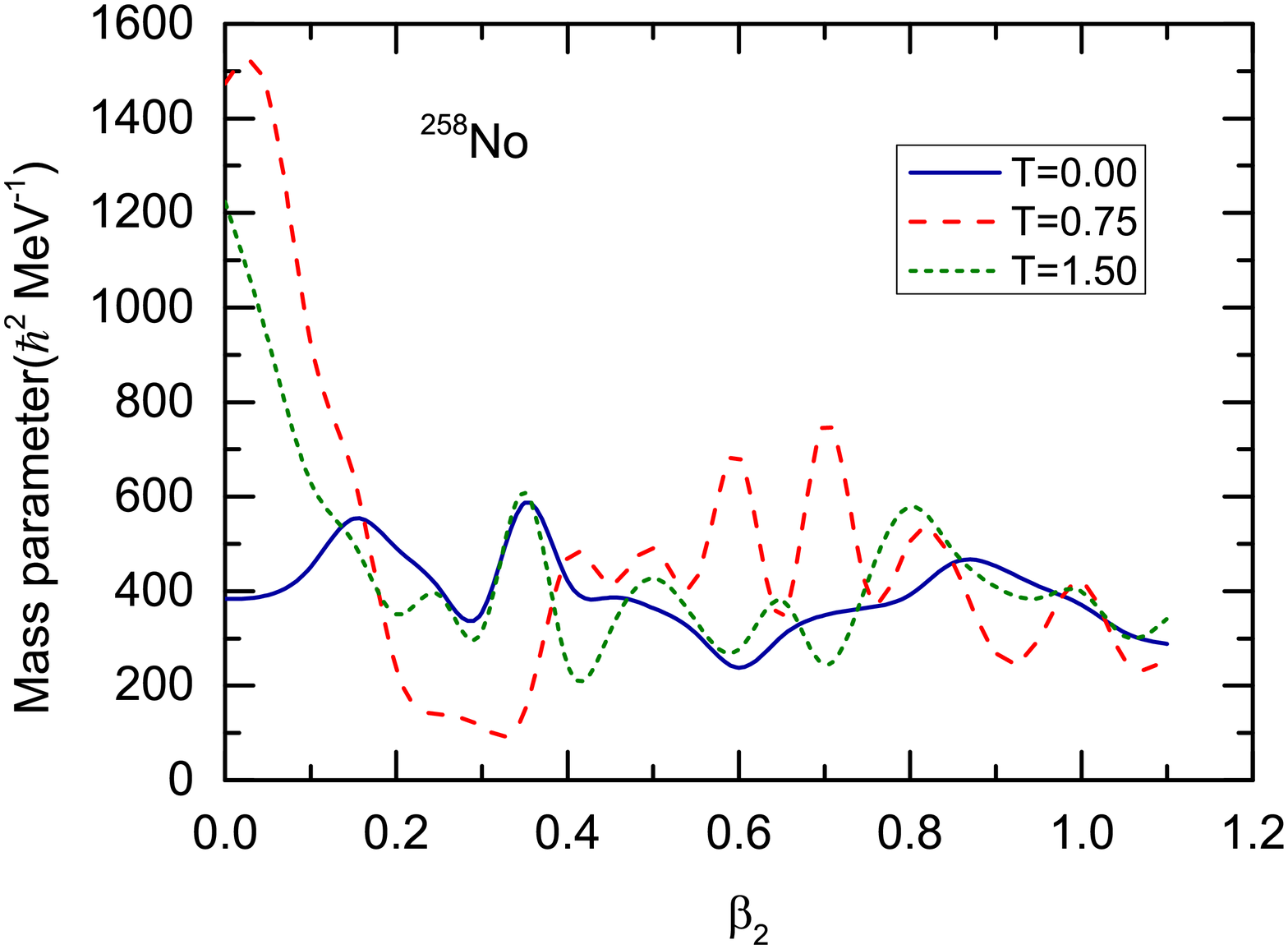}\\
  \caption{(Color online) The mass parameters of $^{258}$No obtained by the temperature-dependent cranking approximation as a function of the deformations. }
  \label{fig-mass}
\end{figure}

The mass parameter is an essential input for the microscopic fission approach.
In this work, the temperature-dependent cranking approximation is employed~\cite{iwamoto1979,baran1994}.
We show in Fig.\ref{fig-mass} the mass parameters of the compound nucleus $^{258}$No for temperatures ranging between 0 and 1.5 MeV.
Based on the discussion of Ref.\cite{bertsch1991}, the mass parameter is inversely proportional to the square of the pairing gap.
As the temperatures increase, the pairing gaps are gradually reduced and finally disappeared at 0.5$\thicksim$0.8 MeV~\cite{khan2007,martin2009}.
Consequently, it can be understood that the mass parameters increase at $T=0.75$ MeV in Fig.\ref{fig-mass} .
At a higher temperature of $T=1.5$ MeV, the disappearance of shell effects\cite{egido2000} leads to reduced mass parameters.

\begin{figure}[t]
  % Requires \usepackage{graphicx}
  \includegraphics[width=0.48\textwidth]{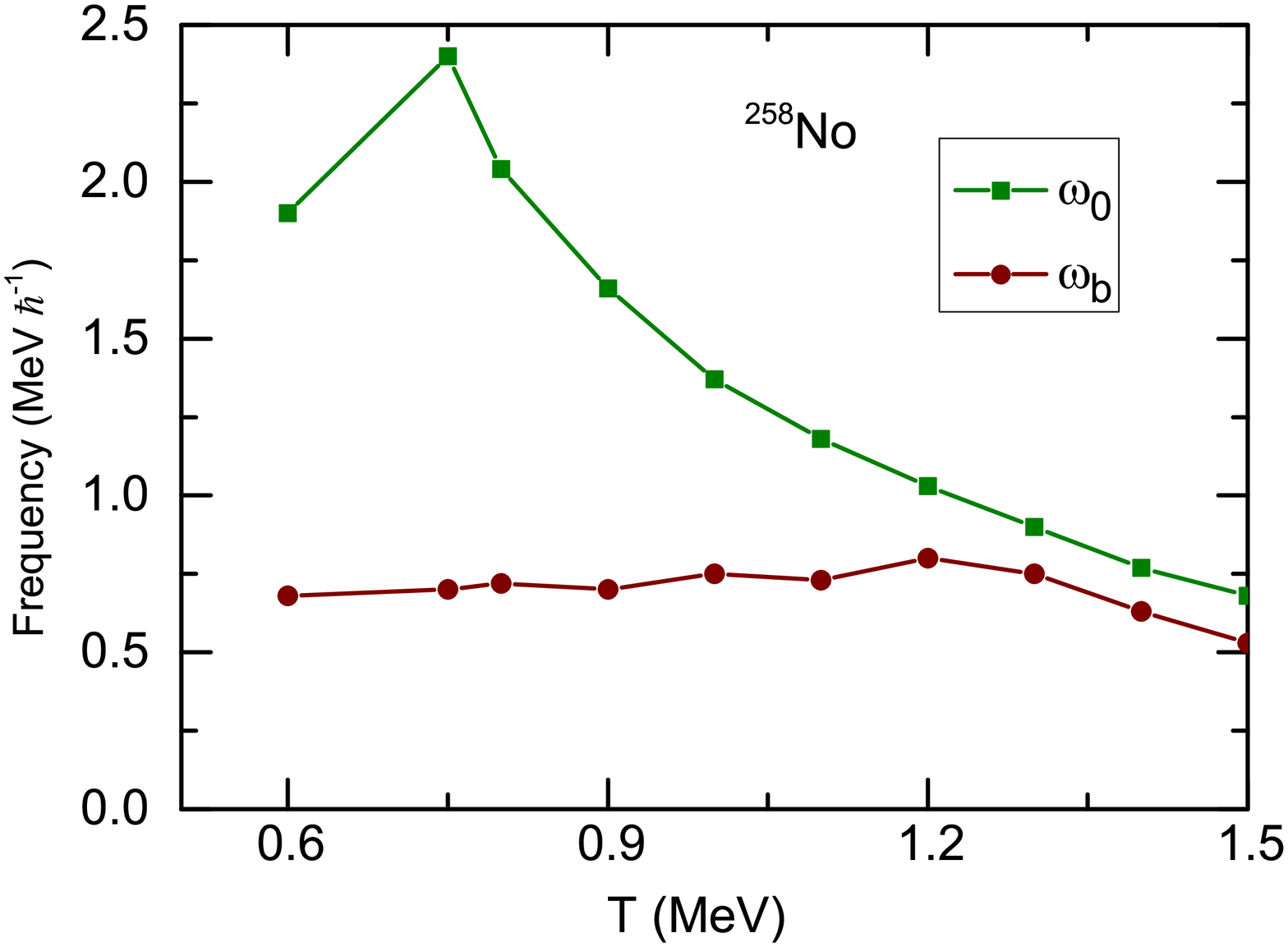}\\
  \caption{(Color online) For $^{258}$No,  the calculated potential curvatures (or frequencies) around the equilibrium point ($\omega_0$)
  and the barrier saddle point ($\omega_b$) as a function of temperature.  }
  \label{fig-omega}
\end{figure}
The fission widths are not only dependent on the heights of the fission barriers but also on the shapes of the barriers.
In Fig.\ref{fig-omega} the curvatures  at the equilibrium point (the potential valley) and the saddle point as a function of temperatures are shown.
We can see that at the equilibrium point,  $\omega_0$ firstly  increases and then decreases.
At the barrier point, $\omega_b$  changes slightly. For different nuclei, the curvatures behavior very differently~\cite{zhu2016}.

\begin{figure}[t]
  % Requires \usepackage{graphicx}
  \includegraphics[width=0.48\textwidth]{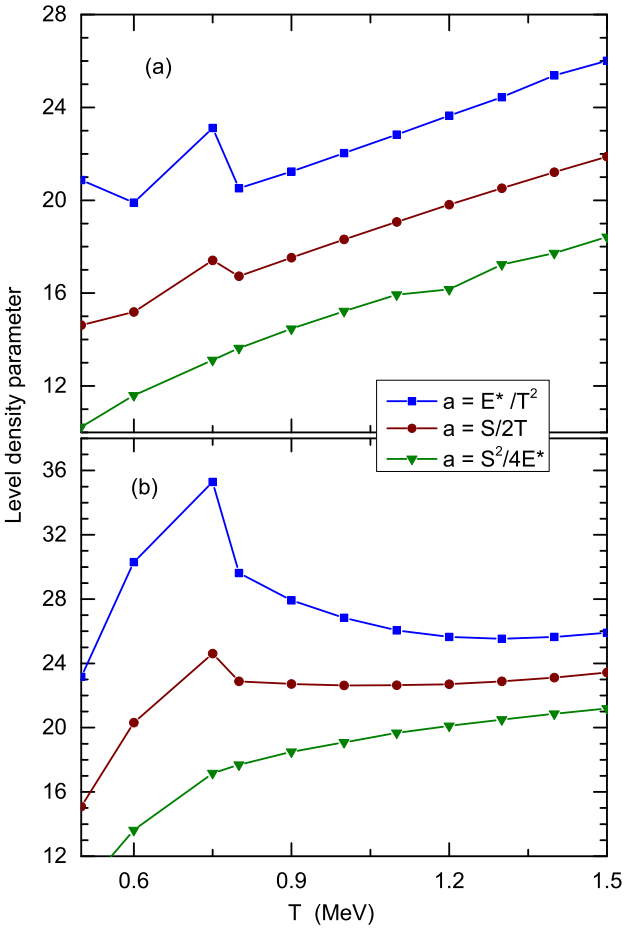}\\
  \caption{(Color online) The level density parameters $a$ of $^{258}$No, calculated by $E^*/T^2$, $S/2T$ and $S^2/4E^*$, respectively,
 at (a) the equilibrium point and  (b) the saddle point, as a function of temperatures.  }
  \label{fig-leveldensity}
\end{figure}
Fig.\ref{fig-leveldensity} displays  the level density parameter $a$ of $^{258}$No calculated by the $E^*/T^2$, $S/2T$ and $S^2/4E^*$~\cite{bonche1984}, respectively .
In panel (a) the level density parameters at equilibrium point ($a_{g.s.}$) with different temperatures are shown.
The results of three different methods have similar trends.
We observe a rapid increase of level density parameters at the temperature $T = 0.75$MeV.
The level density parameters at saddle point ($a_{s.d.}$) from three methods are shown in panel (b).
It can be seen that $a_{s.d.}$ is larger than $a_{g.s.}$  because the level density parameter increases with deformation\cite{pomorski2007}.
The different level density parameters between the ground state and the saddle point have been considered phenomenologically in statistical models.
In the microscopic study, it can be seen that the deformation and temperature dependent level density can be self-consistently taken into account.
The deformation dependence of level densities are related to specific shell structures.
At high excitation energies, the shell effects would disappear and the deformation dependence of level densities would be much reduced.
Indeed, at high temperatures, the level density parameters at the equilibrium point
and the saddle point are close, as shown in Fig.\ref{fig-leveldensity}.

\begin{figure}[t]
  % Requires \usepackage{graphicx}
  \includegraphics[width=0.48\textwidth]{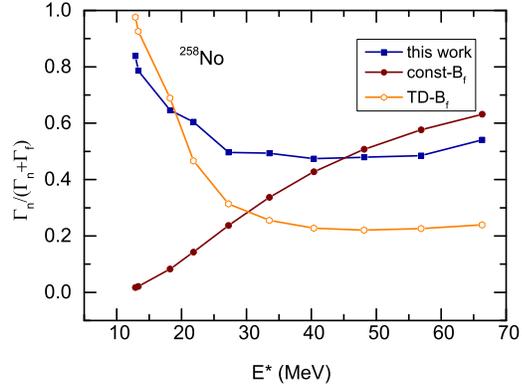}\\
  \caption{(Color online) The calculated $\Gamma_n/\Gamma_{tot}$ as function of  $E^*$ for $^{258}$No with our approach and the statistical model.
  In the statistical model, one adopts a constant fission barrier of 4.53 MeV and one adopts the temperature dependent fission barriers from our calculations.     }
  \label{fig-survival}
\end{figure}
In Ref.~\cite{peterson2009}, the extracted value of $\Gamma_n/\Gamma_{tot}$=$0.840\pm0.050$ for the first-chance fission of $^{258}$No at $E^* = 61$ MeV in the $^{26}${Mg}+$^{232}${Th}.
In Fig.\ref{fig-survival}, the $\Gamma_n/\Gamma_{tot} $  calculated by our approach and the statistical model are plotted versus excitation energies of $^{258}$No.
In our approach for $E^* = 56.9$ MeV, we obtain the neutron emission width $\Gamma_n$=$2.09\times10^{-2}$ MeV, and the fission width $\Gamma_f$=$2.22\times10^{-2}$ MeV.
The final survival probability $ \frac{\Gamma_n}{\Gamma_{n}+\Gamma_{f}}$ is  0.515 that is smaller than experimental value.
 For the statistical model, results with two sets of barrier parameters are shown, for which one adopts a constant barrier height of 4.54 MeV~\cite{peterson2009} and the other is from our temperature-dependent calculations in Fig.~\ref{fig-barrier}.
Fig.\ref{fig-survival} demonstrated that the fission barrier heights have a significant influence on $\Gamma_n/\Gamma_{tot}$.

\begin{figure}[t]
  % Requires \usepackage{graphicx}
  \includegraphics[width=0.4\textwidth]{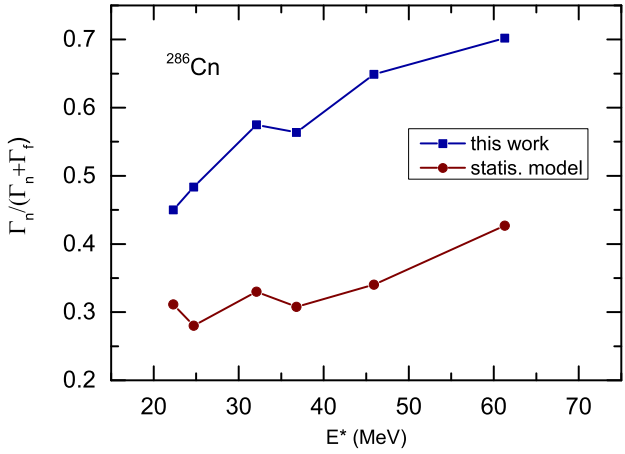}\\
  \caption{(Color online)
  Survival probabilities of $^{286}$Cn obtained by the statistical model  and our microscopic calculations.
  The fission barrier height used in the statistical model is taken from our calculations.  }
  \label{fig-210Po}
\end{figure}

The survival probabilities of $^{286}$Cn have also been studied by the statistical model and our microscopic method, as shown in Fig.\ref{fig-210Po}.
$^{286}$Cn has been studied by the hot-fusion experiment~\cite{itkis2015} and the excitation energy of the compound nuclei are about 40 MeV.
The experimental data of the residual cross section are given for 3$n$ and 4$n$ evaporation channels. There is no direct
measurements for the survival probabilities and a lower limit of $7\times10^{-11}$ is obtained~\cite{itkis2015}.  In this work, we calculate the survival probabilities after the first neutron evaporation, as given
by $\Gamma_n / \Gamma_{tot}$.
Microscopic calculations of survival probabilities after multiple neutron emissions would be extremely time consuming.
In Fig.\ref{fig-210Po}, the survival probabilities are also calculated by the statistical model as described in section(2.4).
The microscopic fission rates are obtained by the Eq.(\ref{espon3}).
In Fig.\ref{fig-210Po}, our results are generally comparable to the statistical model. The microscopic results are
larger than that of the statistical model although the same fission barrier heights are adopted.
This is mainly because the curvatures $\omega_0$ (or frequency) at the equilibrium point of $^{286}$Cn is very small.
The potential energy surface of the compound $^{286}$Cn is very flat around the equilibrium point.
The decreased $\omega_0$ can reduce the fission widths about one order.

\begin{figure}[t]
  % Requires \usepackage{graphicx}
  \includegraphics[width=0.4\textwidth]{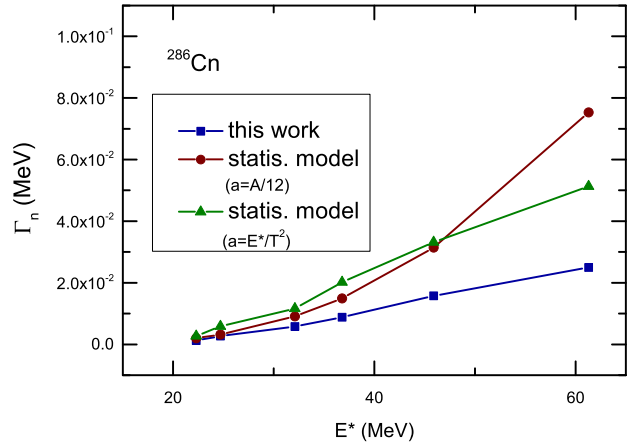}\\
  \caption{(Color online) The neutron emission widths from the  microscopic calculations and the statistical model with the level density parameters of $a=A/12$ and $a = E^*/T^2$ respectively .  }
  \label{fig-neutronemission}
\end{figure}

In order to analyze the difference between the microscopic approach and statistical model, we plot the neutron emission widths of $^{286}$Cn in Fig.\ref{fig-neutronemission}.
It can be seen that results of two methods are comparable within a factor of 3, while the microscopic method underestimates the neutron widths in particular at high energies.
Based on our studies, we can say that the uncertainties of neutron emission rates are smaller than that of fission rates.
In the future, to improve the reliability of microscopic fission theory, we should improve the effective nuclear force~\cite{xiong2016} and also perform multi-dimensional fission calculations.

\section{Summary}\label{summary}
In summary, the microscopic framework for descriptions of survival probabilities of compound superheavy nuclei has been proposed.
Our motivation is to study the microscopic fission rates and neutron emission rates without free parameters, in contrast to
the widely used phenomenological statistical models. The thermal fission rates
are based on the temperature dependent fission barriers from Skyrme-Hartree-Fock+BCS calculations.  We studied the survival probability of
compound nuclei $^{258}$No and $^{286}$Cn. The survival probability of $^{258}$No are comparable to the experimental data.
Generally there are still large uncertainties in fission rates compared to neutron emission rates.
In the future, the microscopic fission rates should be studied in multi-dimensional deformation spaces.

\section{Acknowledgments}

We thank the useful discussions with Profs. G. Adamian, A. Nasirov and Xiaojun Sun.
This work was supported by the National Natural Science Foundation of China under Grants No.11375016, 11522538, 11235001.

\end{document}